\documentstyle[12pt]{article}
\def\beq{\begin{equation}}
\def\eeq{\end{equation}}
\def\bea{\begin{eqnarray}}
\def\eea{\end{eqnarray}}
\def\bq{\begin{quote}}
\def\eq{\end{quote}}

\def\IJMP{{\it Int.J.Mod.Phys.} }

\def\NP{{\it Nucl.Phys.} }
\def\PL{{\it Phys.Lett.} }
\def\PR{{\it Phys.Rev.} }
\def\PRL{{\it Phys.Rev.Lett.} }

\def\ZP{{\it Z.Phys.} }

\def\gappeq{\mathrel{\rlap {\raise.5ex\hbox{$>$}}
{\lower.5ex\hbox{$\sim$}}}}

\def\lappeq{\mathrel{\rlap{\raise.5ex\hbox{$<$}}
{\lower.5ex\hbox{$\sim$}}}}

\begin{document}
\pagestyle{empty}
\begin{flushright}
{CERN-TH.7333/94}
\end{flushright}
\vspace*{5mm}
\begin{center}
{\bf THE ELLIS-JAFFE SUM RULE:} \\
{\bf THE ESTIMATES OF THE NEXT-TO-NEXT-TO-LEADING-ORDER}\\
{\bf QCD CORRECTIONS} \\
\vspace*{1cm}
{\bf A.L. Kataev}$^{*)}$ \\
\vspace{0.3cm}
Theoretical Physics Division, CERN \\
CH - 1211 Geneva 23, Switzerland \\
\vspace*{2cm}
{\bf ABSTRACT} \\ \end{center}
\vspace*{5mm}
\noindent
The procedure of the estimates of the higher-order perturbative QCD
corrections to the physical quantities is generalized to the case when
the
quantities under consideration obey the renormalization group equations
with
the corresponding anomalous dimension functions. This procedure is used
to
estimate the $\alpha^3_s$-corrections to the singlet part of the
Ellis-Jaffe
sum rule for $f = 3$ numbers of flavours.

 \vspace*{5cm}
\noindent
\rule[.1in]{16.5cm}{.002in}

\noindent
$^{*)}$ On leave of absence from Institute for Nuclear Research
of the Russian Academy of Sciences, Moscow 117312,
Russia.
\vspace*{0.5cm}

\begin{flushleft}
CERN-TH.7333/94 \\
July 1994
\end{flushleft}
\vfill\eject

\setcounter{page}{1}
\pagestyle{plain}

The best way of controlling the
theoretical uncertainties of the
perturbative
QCD predictions is the direct analytical or numerical calculation of the
concrete terms in the corresponding perturbative series.
However, after the results of the calculations of the
next-to-next-to-leading
order (NNLO) corrections to the number of physical quantities became
available
\cite{aaa}--\cite{dd}, experimentalists and theoreticians started to be
interested in the effects of still uncalculated higher-order terms. In
the work
of Ref. \cite{ee}, two ``optimal" methods of fixing the renormalization
scheme
ambiguities were used to estimate the next-after-next-to-next-to-leading
order
(NANNLO) corrections to certain renormalization-group invariant
quantities.
These methods were the principle of minimal sensitivity (PMS) \cite{ff}
and the
effective charges approach (ECH) \cite{ggg}, which is known to be
\underline{a
posteriori} identical to the so-called scheme-invariant perturbation
theory
\cite{hh}.
The quantities studied in Ref. \cite{ee} are the $e^+e^-$-annihilation
ratio
$R(s)$, the $\tau$-lepton  decay ratio $R_{\tau}$,
and the Bjorken non-polarized and polarized sum rules.

However, the quantities obeying the renormalization group equations with
anomalous dimension functions were not considered in Refs. \cite{ff},
\cite{ee}.
In this note we will fill in this gap and apply the generalization of
the ideas
used in Ref. \cite{ee} to \underline{estimate} the higher-order
corrections to
the singlet part of the Ellis-Jaffe sum rule (EJSR), recently calculated
at the
$O(\alpha^2_s)$ order \cite{jj}.
We present the concrete $O(\alpha^2_s)$ estimates for $f = 3$ numbers of
flavours in two related forms, namely in the
factorization-scheme-invariant
form and in the form that necessitates the application of an additional
guess
about the value of the unknown four-loop coefficient of the
corresponding
singlet anomalous dimension function.

The experimental measurements \cite{kk} of the structure functions
$g^{p(n)}_1$
of the polarized deep-inelastic lepton-nucleon scattering has stimulated
number of works aimed at a theoretical study of the structure function
$g^{p(n)}_1$ and their first moment, namely the EJSR (see e.g.,
\cite{lll}--\cite{pp}).

The theoretical expression for the EJSR can be presented in the
following form:
\beq
EJSR(Q^2) = \int^1_0 g^{p(n)}_1  (x,Q^2)dx = EJ_{NS}(Q^2)
+ EJ_{SI}(Q^2)~.
\label{1}
\eeq
The non-singlet contribution to this sum rule is defined as
\beq
EJ_{NS}(Q^2) = (1-a-\sum_{i\geq 1} d^{NS}_i a^{i+1})~(\pm
{1\over 12} a_3 +
{1\over
36} a_8)~,
\label{2}
\eeq
where $a = \alpha_s/\pi$, $a_3 = \Delta u - \Delta d$, $a_8 = \Delta u +
\Delta
d - 2\Delta s$ and $\Delta u, \Delta d, \Delta s$ can be interpreted as
the
measure of the polarization of quarks in a nucleon.

The singlet contribution to Eq. (\ref{1}) reads
\beq
EJ_{SI}(Q^2) = C(a) \exp \{\int^{a(Q^2)} {\gamma_{SI}(x)\over\beta (x)}
dx \}
{1\over 9} \Delta \Sigma_{inv}~,
\label{3}
\eeq
where
$$\Delta \Sigma_{inv} = \exp \{- \int^{a(\mu^2)}
{\gamma_{SI}(x)\over\beta (x)} dx \} \Delta \Sigma (\mu^2)
$$
and $\Delta\Sigma = \Delta u + \Delta d + \Delta s$. The first three
coefficients of the QCD $\beta$-function
\beq
\beta (a) = - \sum_{i\geq 0} \beta_i a^{i+2}
\label{4}
\eeq
and of the anomalous dimension of the singlet axial current
\beq
\gamma_{SI} (a) = \sum_{i\geq 0} \gamma_i a^{i+1}
\label{5}
\eeq
are known in the $\overline{\rm MS}$ scheme from the results of
calculations
\cite{qq}, \cite{rr} respectively.
They have the following numerical form:
\bea
\beta_0 &=& ~2.75 - 0.167~f \nonumber \\
\beta_1 &=& ~6.375 - 0.792~f \nonumber \\
\beta_2 &=& 22.320 - 4.369~f + 0.094~f^2
\label{6}
\eea
and
\bea
\gamma_0 &\equiv & 0 \nonumber \\
\gamma_1 &=& -0.5~f \nonumber \\
\gamma_2 &=& -2.458~f + 0.028~f^2~.
\label{7}
\eea
The coefficient $\gamma_3$ is unknown and we will need to somehow fix
its value
in the process of further considerations.

The perturbative expression for the coefficient function
\beq
C(a) = 1 + \sum_{i\geq 1} r_i a^i
\label{8}
\eeq
is explicitly known in the $\overline{\rm MS}$ scheme at the $O(a^2)$
level from the results of Refs.
\cite{jj}, \cite{ss}. The results of direct calculations \cite{jj}
 are
\bea
r_1 &=& -1 \nonumber \\
r_2 &=& -4.583 + 1.162~f~.
\label{9}
\eea
Our aim is to estimate the value of the coefficient $r_3$, in the
$\overline{\rm MS}$ scheme
, thus roughly fixing the uncertainties due to the
lack of
knowledge of the explicitly unknown correction to the singlet part of
the EJSR
at the $O(\alpha^3_s)$ level, already achieved in the direct
calculations of
the related non-singlet contribution \cite{dd}.
This estimate will be obtained
with the help of a procedure that is analogous to the one
of Ref. \cite{ff} which was used in Ref.
\cite{ee} to estimate the $O(\alpha^4_s)$ contribution to the
non-singlet part of the EJSR. As was already demonstrated in
Ref. \cite{ee} this procedure is working quite well at the
$O(\alpha_s^3)$--level. Indeed, the obtained in Ref. \cite{ee}
estimates of the order $O(\alpha_s^3)$-coefficients of the
quantities under consideration are in good agreement with the
results of direct calculations of Refs. \cite{aaa}-\cite{dd}.
This fact stimulates further applications of the methods used.

Let us follow the considerations of Ref. \cite{tt} and define the
renormalization-group-invariant quantity:
\bea
R_{EJ_{SI}} &=& Q^2~{d\ln [EJ_{SI}(Q^2)]\over dQ^2}\nonumber \\
&=& \gamma_{SI}(a) + \beta(a) ~{\partial C(a)/\partial a\over C(a)}
\nonumber \\
&=& d^{SI}_0 a^2 (1+d^{SI}_1 a + d^{SI}_2 a^2)
\label{10}
\eea
where
\bea
d^{SI}_0 &=& -\beta_0 r_1 + \gamma_1 \nonumber \\
d^{SI}_0 d^{SI}_1 &=& -2\beta_0 r_2 + \beta_0 r^2_1- \beta_1r_1 +
\gamma_2
\nonumber \\
d^{SI}_0d^{SI}_2 &=& -3\beta_0r_3 + \gamma_3 -
f(\beta_0,\beta_1,\beta_2,r_1,r_2)
\label{11}
\eea
and
\beq
f(\beta_0,\beta_1,\beta_2,r_1,r_2) = \beta_0r^3_1 - 3\beta_0r_1r_2 +
2\beta_1r_2-\beta_1r^2_1 + \beta_2r_1~.
\label{12}
\eeq
Using the ideas of Refs. \cite{ff},
\cite{uu}, \cite{ee}, one can insert the values
of the
coefficients $d_0,d_1$ and $c_1 = \beta_1/\beta_0$ into the following
expression:
\beq
d^{SI}_0d^{SI}_2 = d^{SI}_0 d^{SI}_1 ({3\over 4} d^{SI}_1 + c_1)~.
\label{13}
\eeq
This can be obtained as the residual term in the re-expansion of the
effective
charge $R_{EJ_{SI}} \equiv d^{SI}_0 a^2_{ECH}$ in terms of the coupling
constant of the initial scheme, in our case the $\overline{\rm MS}$
scheme.
The result of re-expansion of a similar expression obtained
within
the framework of the PMS approach will differ from Eq. (\ref{13}) only
slightly, namely $(d_0^{SI}d_2^{SI})_{PMS}=(d_0^{SI}d_2^{SI})_{ECH}
+(2/6)d_0^{SI}c_1^2$. Therefore, we
 consider the term  of Eq. (13) as the one that simulates the main
contribution of
the non-calculated NNLO correction.

As the result of application of this procedure we get the following
expression
of the coefficient $r_3$:
\beq
r_3 = -{d^{SI}_0d^{SI}_2\over 3\beta_0} + {\gamma_3\over 3\beta_0} -
{f(\beta_0,\beta_1,\beta_2,r_1,r_2)\over 3\beta_0}~,
\label{14}
\eeq
where the anomalous-dimension term $\gamma_3 /(3\beta_0)$ remains
unknown. Note
that this unknown term cancels in the expression for Eq. (\ref{3}) after
taking
into account the corresponding expansion of the anomalous-dimension term
\bea
&&\exp\left\{ \int^a {\gamma_{SI}(x)\over\beta (x)}dx \right\} = 1 -
{\gamma_1\over\beta_0} a\nonumber \\
&&+ \left[ {\gamma^2_1\over\beta^2_0} - {\gamma_2\over\beta_0} +
{\gamma_1\beta_1\over \beta^2_0}\right] {1\over 2} a^2 \nonumber \\
&&+ \left[ -{\gamma^3_1\over 2\beta_0^3} +
\left({\gamma_1\gamma_2\over\beta^2_0} -
{\gamma^2_1\beta_1\over\beta^3_0}\right) {3\over 2} -
\left({\gamma_3\over\beta_0} - {\gamma_1\beta_2\over\beta^2_0} +
{\gamma_1\beta^2_1\over\beta^3_0} -
{\gamma_2\beta_1\over\beta^2_0}\right) \right] {a^3\over 3}~.
\label{15}
\eea
Taking now $f = 3$ numbers of flavours and using Eqs. (\ref{3})-
(\ref{9}) and Eqs. (\ref{11})-(\ref{15})
 we obtain the following
expression for the singlet contribution to the EJSR:
\beq
EJ_{SI}(Q^2) = \bigg[ 1-0.333a - 0.549a^2 -  2 a^3
\bigg] {1\over 9}~~ \Delta \Sigma_{inv}~,
\label{16}
\eeq
where the $O(a)$ and $O(a^2)$ corrections are known from the results of
explicit calculations (see E.g., Ref. \cite{jj}) and the $O(a^3)$
contribution
is our estimate of the value of a correction that is still not
calculated.
Notice once more,
that the possibility to apply the above-described approach for
the
estimates of the $O(\alpha^3_s)$ corrections to the singlet part of the
EJSR is
supported by the good agreement of the results of similar estimates of
the
$O(\alpha^3_s)$ corrections to the corresponding non-singlet
contribution for
$f = 3$ numbers of flavours \cite{ee} with the results of the explicit
calculations of Ref. \cite{dd}.

In order to get a similar estimate for the $Q^2$-dependent normalization
of
$\Delta\Sigma$, namely for the case when $\Delta\Sigma = \Delta\Sigma
(\mu^2 =
Q^2)$, it is necessary to somehow fix the value of the unknown
$\gamma_3$ term
of the singlet anomalous dimension. For $f = 3$, we will use the
following bold
guess-estimate:
\beq
\gamma_3 \approx {\gamma^2_2\over\gamma_1} \approx -34~,
\label{17}
\eeq
which is supported by the detailed consideration \cite{PKK}
of the results of the NNLO calculation of the non-singlet anomalous
dimension for the first four even moments \cite{LRV}.

Using Eq. (\ref{17})  we obtain the bold guess-motivated estimate
of the $O(a^3)$ contribution
to the singlet part of the EJSR in
the case when the $Q^2$ dependence of $\Delta\Sigma$ is specified:
\beq
EJ_{SI}(Q^2) = \left[ 1-a - 1.096a^2 - 3.7a^3\right]~ {1\over
9}~\Delta\Sigma
(Q^2)~.
\label{18}
\eeq
Note, however, that we are unable to present a similar estimate of the
$O(\alpha^3_s)$ correction to the singlet part of the EJSR for $f = 4$
numbers
of flavours. Indeed, for the case of $f = 4$ the value of the
coefficient
$d^{SI}_0$ in Eqs. (\ref{10}) and (\ref{11}) is almost nullified
$(d^{SI}_0
\approx 0$ since $r_1\approx \gamma_1/\beta_0)$. Therefore, it is
impossible to
determine the value of the corresonding coefficient $d^{SI}_1$
from the expression for $d^{SI}_0d^{SI}_1$
presented in Eq.~(\ref{11}). This example demonstrates the limitations
of the
procedure discussed above.

To conclude, we touched the problem of fixing the uncertainties due to
still-uncalculated $O(\alpha^3_s)$ corrections to the singlet
contribution into
the EJSR, for $f = 3$. Combining the obtained estimates with the results
of
available NNLO calculation \cite{dd} of the non-singlet contributions
into the
EJSR and with the corresponding NANNLO estimates \cite{ee}, we arrive at
the
following expressions for the EJSR in the $\overline{\rm MS}$ scheme
, related to $f = 3$:
\bea
\int^1_0 g^{p(n)}_1 (x,Q^2) dx &=& \bigg[ 1-a-3.583a^2 - 20.215a^3 - 130
a^4\bigg] \times \left( \pm {1\over 12} a_3 + {1\over 36} a_8\right)
\nonumber \\
&& +\bigg[ 1-0.333a - 0.549a^2 - 2a^3 \bigg] ~{1\over 9} ~
\Delta\Sigma_{inv}~,
\label{19}
\eea
or
\bea
\int^1_0 g^{p(n)}_1 (x,Q^2) dx &=& \bigg[ 1-a-3.583a^2 - 20.215a^3 - 130
a^4\bigg] \times \left( \pm {1\over 12} a_3 + {1\over 36} a_8\right)
\nonumber \\
&& +\bigg[ 1-a-1.096a^2 - 3.7a^3 \bigg] ~{1\over 9} ~
\Delta\Sigma (Q^2)~.
\label{20}
\eea

It can be seen that the perturbative contributions to the singlet part
of the
EJSR, including the $O(\alpha^3_s)$ term that we estimated, are
negative. They
are significantly smaller than the coefficients of the perturbative
series of
the non-singlet part, which include the results of the concrete
$O(\alpha^3_s)$ calculations \cite{dd} and the estimates of the
$O(\alpha^4_s)$ terms \cite{ee}, that are in qualitative agreement
with the results of application of the Pad\'e resummation technique
\cite{sls}.
  The concrete physical applications
of the results obtained are discussed in Ref. \cite{pp}.

\vspace*{1cm}
\noindent
{\bf Acknowledgements}

We are grateful to J. Ellis and M. Karliner for the interest in the work
done
in collaboration with V. Starshenko \cite{ee} and for useful
discussions,
which stimulated the considerations presented in this note.
It is also the pleasure to thank E. Hughes and W.J. Stirling for the
interest in the results of these considerations.

\end{document}